\documentclass{Interspeech}

\interspeechcameraready

\title{BR-ASR: Efficient and Scalable Bias Retrieval Framework for \\ Contextual Biasing ASR in Speech LLM}

\author[affiliation={1}]{Xun}{Gong}
\author[affiliation={2}]{Anqi}{Lv}
\author[affiliation={2}]{Zhiming}{Wang}
\author[affiliation={2}]{Huijia}{Zhu}
\author[affiliation={1*}]{Yanmin}{Qian}

\affiliation{Auditory Cognition and Computational Acoustics Lab, MoE Key Lab of AI}{AI Institute\\ School of Computer Science, Shanghai Jiao Tong University}{Shanghai, China}
\affiliation{AntGroup}{Shanghai}{China}

\email{\{gongxun,yanminqian\}@sjtu.edu.cn,\{lyuanqi.laq,zhiming.wzm,huijia.zhj\}@antgroup.com}
\keywords{contextual bias, speech retrieval, contrastive learning, speech large language model, speech recognition}

\usepackage{comment}
\usepackage{multirow}
\usepackage{bbding}
\usepackage{booktabs}
\usepackage{makecell}
\usepackage{float}
\usepackage{graphicx}

\usepackage{setspace} 
\usepackage[labelfont=bf]{caption} 
\newcommand{\bias}{\mathcal{B}}
\newcommand{\bX}{\mathbf{X}}
\newcommand{\by}{\mathbf{y}}
\newcommand{\bb}{\mathbf{b}}
\newcommand{\bH}{\mathbf{H}}
\newcommand{\recallh}{$\text{Recall}_{\text{H}}$}
\newcommand{\recallb}{$\text{Recall}_{\text{B}}$}

\begin{document}

\maketitle

\begin{abstract}
While speech large language models (SpeechLLMs) have advanced standard automatic speech recognition (ASR), contextual biasing for named entities and rare words remains challenging, especially at scale.
To address this, we propose \textbf{BR-ASR}: a \textbf{B}ias \textbf{R}etrieval framework for large-scale contextual biasing~(up to 200k entries) via two innovations:
(1) speech-and-bias contrastive learning to retrieve semantically relevant candidates;
(2) dynamic curriculum learning that mitigates homophone confusion which negatively impacts the final performance.
The is a general framework that allows seamless integration of the retrieved candidates into diverse ASR systems without fine-tuning.
Experiments on LibriSpeech test-clean/-other achieve state-of-the-art (SOTA) biased word error rates (B-WER) of 2.8\%/7.1\% with 2000 bias words, delivering 45\% relative improvement over prior methods.
BR-ASR also demonstrates high scalability: when expanding the bias list to 200k where traditional methods generally fail, it induces only 0.3\,/\,2.9\% absolute WER\,/\,B-WER degradation with a \textbf{99.99\%} pruning rate and only 20ms latency per query on test-other.
\end{abstract}

\section{Introduction}

The integration of speech processing into large language models (SpeechLLMs) has advanced speech tasks like automatic speech recognition (ASR), achieving state-of-the-art performance in general domains~\cite{ji2024wavchat}, making them more applicable in real-world scenarios such as voice assistants.
However, challenges arise in accurately recognizing and understanding bias words~(also known as hotwords), which are rare terms such as names, locations, or uncommon slang~\cite{pundak2018clas,yang2023promptasr,wang2023text,salemi2023llm_personal}.
To address this issue, many contextual biasing ASR systems are developed to improve the bias word recognition accuracy without dropping the overall ASR performance~\cite{chen2024prompt_salm,yang2023promptasr,yang2024ctc}.

A straightforward approach is to post-process ASR hypotheses with bias words by leveraging text-only LLMs' linguistic reasoning ability~\cite{song2023contextual,asano2025contextualasrerrorhandling}. 
However, these text-only approaches inherently lack speech modality awareness (e.g., accent, prosody), resulting in suboptimal performance compared to speech-grounded methods~\cite{lakomkin2024prompt_llm_meta}.
This limitation has driven recent efforts to develop prompt-based contextual biasing through multimodal SpeechLLMs~\cite{gong2024contextual,lakomkin2024prompt_llm_meta,chen2024prompt_salm,hou2025contextual,li2023prompt_llm_domain,gong2025contextual}, where acoustic-textual alignment enhances robustness and recognition accuracy.
In these methods, the prompt typically comprises a bias word list or domain-specific instructions.
However, three critical limitations persist:
(1) Excessive prompt length induces LLM hallucinations, leading to unrealiable ASR outputs;
(2) Increased computational complexity from lengthy prompts degrades inference efficiency.
(3) Bias list is limited to dozens of entries, severely limiting the scalability.

To scale contextual biasing in ASR, recent approaches adopt retrieval-augmented generation (RAG) paradigms from LLMs \cite{gao2024rag} integrating external knowledge database.
Methods like DOC-RAG \cite{mathur2024doc}, LA-RAG \cite{li2024rag_asr}, and GEC-RAG \cite{robatian2025gec_rag} use LLMs for ASR hypotheses post-processing.
However, two more challenges in scalability and efficiency exist:
(1) Domain-specific knowledge bases lack fine-graind alignment with exact bias words, limiting the bias recognition accuracy.
(2) Retrieval mechanisms rely on rough cross-modal similarity withou deep speech-text modality alignment.

Current fine-grained biasing methods exploit localized acoustic or semantic patterns.
Phoneme-RAG~\cite{lei2024phone_rag} employs multi-stage pipelines, where entities are detected at the first stage and then retrieved from the whole database and fed to the LLM.
CTC-filter~\cite{yang2024ctc} utilizes CTC N-best hypotheses to filter relevant bias words.
However, Phoneme-RAG's two-stage inference incurs significant latency with limited performance improvement.
CTC-filter suffers from irreversible errors when correct candidates are absent from CTC N-best hypotheses.
While achieving better accuracy, both methods scale poorly beyond 10k entries and lack dynamic database integration, fundamentally limiting their industrial applicability.

Recent attention-based advancements focus on frame-level features, and extend the bias number to 40k–100k entries through Dual-NAM~\cite{huang24retrieval_asr}, and VQ-RAG \cite{flemotomos2024vq_retrieval}.
However, the performance still degrades linearly beyond 10k entries with excessive computational overhead.
This underscores the unresolved trade-off between bias list size, accuracy, and inference speed in large-scale contextual biasing ASR.

\begin{figure*}[htbp]
    \centering
    \includegraphics[width=\linewidth]{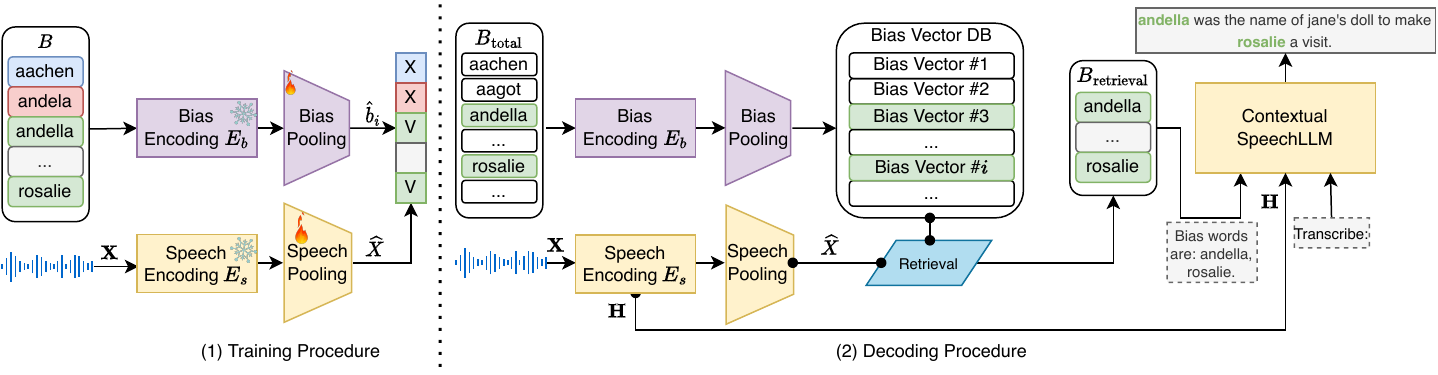}
    \caption{The proposed Bias Retrieval Framework with training~(1) and decoding~(2) procedures for Contextual SpeechLLM.}
    \label{fig:main}
    \vspace{-1.5em}
\end{figure*}

To bridge these gaps, we propose \textbf{BR-ASR}: an efficient and scalable \textbf{B}ias \textbf{R}etrieval framework that fundamentally mitigates scalability bottlenecks for contextual ASR.
Our approach begins with a contrastive learning mechanism that optimizes speech-bias matching via two different bias encoding strategies, enabling effective retrieval between speech inputs and large-scale bias entries.
Besides, speech encoding and bias encoding procedures are inherited from SpeechLLMs, simplifying the training procedure.
Building on this foundation, we design a dynamic curriculum scheduler that automatically modulates hard negative sampling ratios during training, specifically homophone confusion, a critical bottleneck in large-vocabulary biasing.
The proposed BR-ASR framework is evaluated on the LibriSpeech benchmark with Rare5k~\cite{le2021contextualized}.
The proposed framework achieves 93\%\,/\,91\% retrieval accuracy on test-clean\,/\,-other subsets and delivers state-of-the-art B-WER of 2.8\%\,/\,7.1\% over all previous methods.
As the number of bias words scales up to 200k, it preserves a stable performance with only 0.3\%\,/\,2.9\% absolute WER\,/\,B-WER degradation on the test-other subset, resolving the long-tail coverage issues.
With the help of FAISS~\cite{johnson2019faiss}, the retrieval latency is only 20ms when querying 200k entries on GPU.
Crucially, our framework demonstrates strong generalization: when integrated with third-party ASR systems, it consistently improves rare-word recognition without architectural modifications.

\section{Methodology}

\subsection{Revisiting Contextual Biasing in SpeechLLM-based ASR Systems}  

The contextual ASR~(C-ASR) can be formally defined through three core components:  
(1) the input audio spectrogram $\bX \in \mathbb{R}^{M \times T}$, where $M$ denotes mel-frequency bins and $T$ temporal frames, (2) a predefined bias list $\bias = \{b_1, b_2, \dots, b_N\}$ containing $N$ domain-specific terms or rare entities (where the oracle bias words is defined as $\bias_{\text{sentence}}$, s.t. $|\bias_{\text{sentence}}| \le N$), and (3) the target transcription $\by$.
The objective is modeled as an autoregressive conditional probability:
\begin{align}
P(y_{i+1} | \bX, \by_{1:i}, \bias) = \text{C-ASR} (\phi(\bX), \by_{1:i}, \psi(\bias)), \label{eq:casr}
\end{align}
where SpeechLLM implementations involve two critical processes:
(1) Speech encoding $\phi(\cdot)$: Audio inputs are encoded through $\phi(\cdot)$ into latent representations $\bH$, which serve as speech latent features for the SpeechLLM.
(2) Bias integration: The bias list $\bias$ is incorporated via $\psi(\cdot)$, implemented either as textual prompts (e.g., ``Related bias words are $b_1, b_2, \cdots$''.~\cite{lakomkin2024prompt_llm_meta,gong2023longfnt,chen2024prompt_salm} or through dedicated bias encoders~\cite{gong2023factorized,gong2024advanced}.

\subsection{Bias Retrieval Framework} 
\label{sec:bias_retrieval}

\subsubsection{Contrastive Learning for Speech-and-Bias Alignment}
\label{sec:clap}

Direct similarity computation between textual bias words and raw speech signals proves ineffective due to modality discrepancies and granularity mismatches.
To address this, we redesign contrastive language-audio pretraining (CLAP)~\cite{elizalde2023clap} for speech-and-bias alignment, as illustrated in Fig.~\ref{fig:main}~(1).

Our modified training procedure introduces two encoding network, speech encoding network $E_s$, and bias encoding network $E_b$, the pre-training procedure can be formulated as:
\begin{align}
\hat{\bX} &= \text{Pool} \cdot E_s(\bX), \quad \hat{\bb} = E_b(b), \label{eq:clap:proj} \\
\mathcal{L} &= 0.5 * (l_a(S) + l_b(S)), \text{where } S = \tau (\hat{\bX} \cdot \hat{\bb}^\intercal), \label{eq:clap:loss}
\end{align}
where $\hat{\bX} \in \mathbb{R}^{d}, \hat{\bb} \in \mathbb{R}^{d}$ are encoded speech/bias features of dimension $d$, and $\text{Pool}(\cdot)$ is a special network to make speech features the same length as the bias features.
For the objective loss, $S$ is the dot-product similarity score, $\tau$ is a learnable temperature and $l_k$ is the sum of logits along text and audio axes respectively.
Following CLAP~\cite{elizalde2023clap}, let $N_{neg}$ be the number of negative samples during training, we randomly select $N_{neg}$ bias words from the bias database $\mathcal{B}_{\text{total}} \setminus \mathcal{B}_{\text{sentence}}$, where $\mathcal{B}_{\text{sentence}}$ is the oracle bias list for each sentence.

\subsubsection{Frozen Encoding Priors}
\label{ssec:encoding}

To prevent catastrophic forgetting and ensure feature consistency, we fully freeze speech/bias encoding network $E_s , E_b$ shown in Fig.~\ref{fig:main}~(1).

\noindent \textbf{Frozen Encoding Priors}:
As a well-trained speech encoder is available in SpeechLLMs, we reuse the speech encoder $E_s(\cdot) = \phi(\cdot)$, and thereby the dimension mismatch between speech latent and bias latent can be simplified into two parameter-efficient ways for different bias encoding network:
\begin{itemize}
\item \textit{AcousticBias}:
The bias word is firstly synthesized to audio, then the same encoding network is applied as $E_b (b) = E_s (\text{TTS}(b))$.
This approach simplifies cross-modal alignment by projecting both speech inputs and bias terms into the same acoustic space.

\item \textit{TextualBias}:
We also directly employ the same LLM backend to extract textual bias features, i.e. $E_b(b) = \text{LLM} (b)$.
This preserves linguistic semantics while maintaining model interoperability.
For this setup, $E_s(\cdot) = \text{Projection} \cdot \text{Speech-Encoder} (\cdot)$.
\end{itemize}

\noindent \textbf{Pooling Strategies for speech representation}
To address length mismatch between speech latent features and bias latent features, we implement three strategies to pool the frame-level speech features $\bH$ from $D\times T'$ to the pooled feature $\hat{\bX}$:
\begin{itemize}
\item \textit{Adaptive Average Pooling}: The same pooling as CLAP~\cite{elizalde2023clap}.
\item \textit{Attention Pooling}: Let $d$ be the dimension of attention calculations, the output is $\hat{\bX} = \text{Linear} \cdot \sum_{T'} (\text{Attention} (\bH))$.
\end{itemize}

\subsubsection{Dynamic Curriculum Learning for Homophone Debias}
\label{sec:curriculum}

To mitigate homophone confusion (e.g., "there" vs. "their"), we develop a phoneme-aware dynamic curriculum learning (DCL) mechanism~\cite{gong2022knowledge,gong2021speaker,qian2022layer}.
Firstly, we use grapheme-to-phoneme (G2P) to generate phoneme sequence and and construct homophone set $\mathcal{H}_i$ for bias word $b_i$:
\begin{align}
\mathcal{H}_i = \{ h_j | \text{Lev}(p_i, p_j) \leq \theta, h_j \in \mathcal{B} \setminus \{b_i\} \},
\end{align}
where Lev is Levenshtein edit distance and $\theta=2$ is the phoneme mismatch threshold.
Then we dynamically adjust the homophone sampling ratio through sigmoidal scheduling:
\begin{align}
\alpha_{n} = \alpha_{\min} + (\alpha_{\max} - \alpha_{\min}) \cdot \left(\frac{2}{1+e^{-\gamma n}} - 1\right),
\end{align}
where $n$ denotes the training steps, and $\gamma$ is a controllable parameter.
To further disperse homophone in training, we apply an additional regularization term in the objective function:
\begin{align}
\mathcal{L}_{\text{total}} = \mathcal{L} + 
\lambda \sum_{h_j \in \mathcal{H}_i} 
\Vert b_i - h_j \Vert^2_2, \label{eq:loss_reg}
\end{align}
where $b_i$ is the bias word embedding and $h_j$ is word embedding from one of the homophones of $b_i$.

\section{Experiments}

\subsection{Experimental Setup}

\subsubsection{Dataset and Metrics}

Experiments were conducted using LibriSpeech~\cite{panayotov2015librispeech} for both training and evaluation, where bias words are marked by Rare5k~\cite{le2021contextualized}. Rare5k identifies the top 5k most frequent words as common, with the remaining 209.2k words classified as rare in the LibriSpeech train-960h set.

Following the same setup in \cite{le2021contextualized}, we evaluate word error rate~(WER)~(\%) and biased WER~(B-WER)~(\%), where B-WER is calculated over bias words on test-clean/-other subsets.

We further consider two extra evaluation dimensions to directly evaluate the retrieval performance by special recall variants\footnote{
Recall@X denotes the average retrieval number $\overline{N}_{\text{retrieval}}$ when the Recall = X\% over dataset.
Recall\#X denotes the recall rate~(\%) when $|\bias_{\text{retrieval}}| = X$ for each sentence.
}.
(1) \textbf{Bias Retrieval Recall}:
the recall rate~(\%)~($\uparrow$) of bias list over all sentences, i.e. $\text{Recall}_{\text{B}} = \frac{\sum |\bias_{\text{sentence}}|}{\sum |\bias_{\text{retrieval}}|}$, where $\bias_{\text{retrieval}}$ denotes the retrieved bias list.
(2) \textbf{Homophone Retrieval Recall}:
the recall rate~(\%)~($\downarrow$) of all homophones over all sentences, i.e. $\text{Recall}_{\text{H}} = \frac{\sum |\mathcal{H}_{\text{sentence}}|}{\sum |\bias_{\text{retrieval}}|}$\footnote{$\mathcal{H}_{\text{sentence}} = \{ h | h \in \bigcup\limits_i \mathcal{H}_i, \forall i \in \bias_{\text{sentence}}\}$}.

\subsubsection{Model setups and hyper-parameter selection}

Two different SpeechLLMs are selected as the contextual ASR backend, Qwen-Audio~\cite{chu2023qwenaudio} and SLAM-ASR~\cite{ma2024slam_asr}.
For Contextual Qwen-Audio, we follow two different contextual methods, i.e., bias-prompt~(Prompt-QwenAudio) and bias-encoder~(BiasEncoder-QwenAudio) of \cite{gong2024contextual}. 
The SpeechLLM backend has been changed to the unchat version\footnote{https://huggingface.co/Qwen/Qwen-Audio}.
The Prompt-QwenAudio is trained using the bias size between 20$\sim$80.
The BiasEncoder-QwenAudio is trained using the bias size between 50$\sim$1000, and the bias encoder is initialized from \texttt{gte-large-en-v1.5}\footnote{https://huggingface.co/Alibaba-NLP/gte-large-en-v1.5}.
For the Contextual SLAM-ASR, we reuse the pre-trained Prompt-SLAM-ASR \cite{yang2024ctc}, where the training bias size is 5 on average.

For the proposed Bias Retrieval, hyper-parameters are set as follows.
For AcousticBias defined in Section~\ref{ssec:encoding}, we firstly synthesized all bias words from the database $\bias_{\text{total}}$ using \texttt{edge-tts}\footnote{https://github.com/rany2/edge-tts} with the default voice ``en-US-AnaNeural''.
Then both audio and bias encodings are fed to the same speech encoder used in SpeechLLMs.
For TextualBias defined in Section~\ref{ssec:encoding}, the bias encoding reuses the textual LLM part from SpeechLLMs, and the audio encoding is done by the speech encoder with projection.
Note that there is a dimension mismatch between these encodings, so the same projection inherited from SpeechLLMs is reused.
For the contrastive learning proposed in Section~\ref{sec:clap}, the temperature $\tau$ is initialized to 0.007 and clipped by at most 100 in Eq.~\ref{eq:clap:loss}.
During training, we use AdamW with a maximum learning rate of $1e^{-4}$ warmup ratio of 0.05 and train the network for 10 epochs with 960h data.
As for the dynamic curriculum learning in Section~\ref{sec:curriculum}, we set the homophone distance threshold $\theta$ to 2, and the parameters as $\alpha_{\min}=0.01,\alpha_{\max}=0.5, \gamma = 0.05$.
The regularization loss parameter $\lambda$ in Eq.~\ref{eq:loss_reg} is set to $0.1$.

Finally, for the scalable retrieval, we use FAISS~\cite{johnson2019faiss}, where vectors are pre-normalized and then retrieved with METRIC\_INNER\_PRODUCT. 
The retrieval threshold is set to 50, 50, and 10 for Prompt-QwenAudio, BiasEncoder-QwenAudio and Prompt-SLAM-ASR, respectively.

\subsection{Main Results of Bias Retrieval Method}

{
\setlength{\tabcolsep}{4pt}
\begin{table*}[ht]
\centering
\caption{Performance of the contextual ASR baselines and our proposed \textbf{Bias Retrieval} method on LibriSpeech test-clean/-other sets.}
\label{tab:main}
\resizebox{0.96\textwidth}{!}{%
\begin{tabular}{l|cc|cc|cc|cc|cc}
\toprule
\multirow{2}{*}{Contextual ASR Model}
& \multicolumn{2}{c|}{$N=|\bias_{sentence}|$} & \multicolumn{2}{c}{$N=100$} & \multicolumn{2}{|c}{$N=500$} & \multicolumn{2}{|c}{$N=1000$} & \multicolumn{2}{|c}{$N=2000$} \\
& WER & B-WER & WER & B-WER & WER & B-WER & WER & B-WER & WER & B-WER \\
\midrule \midrule
DB-NNLM~\cite{le2021contextualized} & - & - & 2.0/5.9 & 5.7/14.1 & 2.1/6.1 & 6.2/15.1 & 2.1/6.4 & 6.7/17.2 & 2.3/6.6 & 7.3/18.9 \\
USTR-CT~\cite{jin2023bytedance} & - & - & 2.1/5.4 & \textbf{2.0/4.4} & 2.1/5.6 & \textbf{2.2/5.6} & 2.2/5.8 & \textbf{2.5}/6.3 & 2.2/5.8 & 3.0/7.6 \\ 
\midrule \midrule
CB-QwenAudio~\cite{gong2024contextual} & - & - & 1.6/3.8 & 5.5/13.5 & 1.9/3.9 & 6.0/14.2 & - & - & - & - \\
Prompt-QwenAudio       & 1.1/2.6 & 1.2/3.7 & 1.3/3.5 & 2.4/6.4 & 22.4/27.8 & 32.1/39.1 & X & X & X & X \\ 
\quad + \textbf{Bias Retrieval} & 1.1/2.6 & 1.2/3.7 & 1.2/2.7 & {2.3/6.1} & 1.2/2.7 & {2.4/6.3} & 1.2/2.7 & \textbf{2.6/6.6} & 1.2/2.8 & \textbf{2.8/7.1} \\
BiasEncoder-QwenAudio  & 1.4/3.3 & 3.8/9.0 & 1.5/3.6 & 5.0/13.0 & 1.6/3.7 & 5.6/13.8 & 1.7/3.9 & 6.2/14.7 & 1.8/4.1 & 7.4/18.1 \\
\quad + \textbf{Bias Retrieval} & 1.4/3.3 & 3.8/9.0 & 1.5/3.4 & 4.0/9.5 & 1.5/3.5 & 4.2/9.6 & 1.5/3.6 & 4.3/9.8 & 1.5/3.6 & 4.3/9.9 \\
\midrule \midrule
Prompt-SLAM-ASR~\cite{yang2024ctc} & 1.1/2.7 & 2.8/6.0 & 7.4/17.9 & 24.8/44.1 & X & X & X & X & X & X \\
\quad + CTC-Filter~\cite{yang2024ctc} & 1.1/2.7 & 2.8/6.0 & 1.3/2.7 & 3.7/8.0 & 1.3/3.0 & 3.9/9.0 & 1.3/3.0 & 4.2/9.3 & 1.4/3.2 & 4.4/10.0 \\
\quad + \textbf{Bias Retrieval} & 1.1/2.7 & 2.8/6.0 & 1.3/2.7 & 3.8/8.3 & 1.3/3.0 & 4.0/8.7 & 1.3/3.1 & 4.1/9.0 & 1.4/3.1 & 4.2/9.3 \\
\bottomrule
\end{tabular}%
}
\end{table*}
}

{
\setlength{\tabcolsep}{4pt}
\begin{table*}[ht]
\centering
\caption{Ablation Studies on LibriSpeech test-other set with $N=2000~(2k)$ of Prompt-QwenAudio as Contextual ASR system.}
\label{tab:ablation}
\setstretch{0.92}
\resizebox{0.7\textwidth}{!}{%
\begin{tabular}{c|c|c|c|c|c|c|c}
\toprule
\makecell{Bias \\ Modality} & \makecell{Speech \\ Pooling} & \makecell{Curriculum \\ Learning} & \recallb@99 ($\uparrow$) & \recallb\#50 ($\uparrow$) & \recallh\#50 ($\downarrow$) & WER & B-WER \\
\midrule
\multirow{3}{*}{Acoustic}
& AdaAvgPool & \XSolidBrush & 54.1 & 98.1\% & 98.7\% & 3.3 & 12.4 \\
& AttnPool   & \XSolidBrush & \textbf{32.8} & \textbf{99.9\%} & 99.9\% & 3.0 & 8.6 \\
& AttnPool   & \Checkmark   & 42.2 & 99.7\% & 69.3\% & 2.9 & 7.4 \\
\midrule
\multirow{3}{*}{Textual}
& AdaAvgPool & \XSolidBrush & 58.5 & 97.3\% & 94.9\% & 3.1    & 10.3 \\
& AttnPool   & \XSolidBrush & 38.7 & 99.5\% & 91.3\% & 3.0    & 7.9 \\
& AttnPool   & \Checkmark   & 47.3 & 99.1\% & \textbf{58.4\%} & \textbf{2.8} & \textbf{7.1} \\
\bottomrule
\end{tabular}%
}
\vspace{-1.2em}
\end{table*}
}

Table \ref{tab:main} compares the performance of contextual ASR models on LibriSpeech test sets.
We firstly explore the performance on QwenAudio variants.
The baseline QwenAudio achieves WER=$2.0\%/4.2\%$ and B-WER=$8.4\%/18.4\%$ on test-clean/-other, respectively.
Its fine-tuned version obtains WER=$1.7\%/3.8\%$ and B-WER=$8.3\%/18.5\%$, indicating that pure domain adaptation fails to improve bias recognition accuracy.
We reproduce two implementations based on \cite{gong2024contextual}, both surpassing the original CB-QwenAudio.
Results show that Prompt-QwenAudio suffers catastrophic failure (at $N\ge100$) due to LLM hallucinations, and unreliable results for $N\ge1000$ due to exceeded context limits.
In contrast, BiasEncoder-QwenAudio obtains better results when $N=1000,2000$.

With the proposed Bias Retrieval framework, both achieve systematic improvements: Prompt-QwenAudio+BR reduces B-WER by over 80\% absolutely when N=500 (6.3\% vs 32.1\%), and continues working well when N=1000/2000.
BiasEncoder-QwenAudio+Bias-Retrieval keeps the B-WER below 10.2\% for nearly all N values, which is attributed to its ability to maintain better stability through the secondary text processing that effectively disambiguates homophones.

Compared to conventional approaches, previous SOTA, USTR-CT exhibits a B-WER increase from 4.4\% to 7.6\%.
In contrast, our method demonstrates superior robustness while maintaining a stable B-WER (7.1\% for Prompt+BR and 10.2\% for BiasEncoder+BR), by modeling speech-and-bias correlation.
This also represents 17.1\% relative improvement over USTR-CT's best baseline performance when N=2000.
Our framework ultimately achieves breakthrough performance with B-WER=2.8\%/7.1\% when N=2000, establishing a new SOTA for large-scale contextual ASR.

In addition the results on the SLAM-ASR system also validate our framework's generalization capability, showing similar trends in preserving the B-WER performance.

\subsection{Ablation Studies}

Based on the ablation study results in Table \ref{tab:ablation}, we systematically analyze key designs of the bias retrieval framework: 

\textbf{Speech Pooling Strategy}: It shows that AcousticBias with AdaAvgPool does not work effectively as \recallh@99 $>$ 50. 
In contrast, AttnPool provides a better pooling, focusing on the semantic information and increasing the retrieval efficiency: \recallb@99 reduced greatly from 54.1 to 38.7. This also implies that almost all bias words can be successfully retrieved within top-50~(\recallb\#50).
With AttnPool, the B-WER improves by 32\% relatively. 
 
\textbf{Dynamic Curriculum Learning}:
Although \recallh\#50 is high, we found that there is a huge gap between $N = |\bias_{\text{sentence}}|$~(In Table~\ref{tab:main}, B-WER=3.7\% on test-other) and $N=|\bias_{\text{retrieval-top-50}}|$.
We observed that homophones are mistakenly recognized~(i.e., homophones confusion), as \recallh\#50 is also high for both AdaAvgPool and AttnPool.
The proposed dynamic curriculum learning method greatly suppresses homophone errors.
When activated, the AcousticBias reduces \recallh\#50 from 99.9\% to 69.3\%, with another 23\% relative reduction on B-WER on test-other.

\textbf{Bias Encoding Modality}:
To further mitigate homophone errors during retrieval, we adopt TextualBias, as textual inputs avoid the domain mismatch between synthesized and real speech, and are more robust to OOV words.
Compared to AcousticBias, TextualBias achieves competitive retrieval performance and lowers the homophone recall rate (\recallh\#50 reduced by 10\%), with a 0.3\% absolute reduction in B-WER.

\subsection{Scalability Analysis}

\begin{figure}[t]
\centering 
\includegraphics[width=\linewidth]{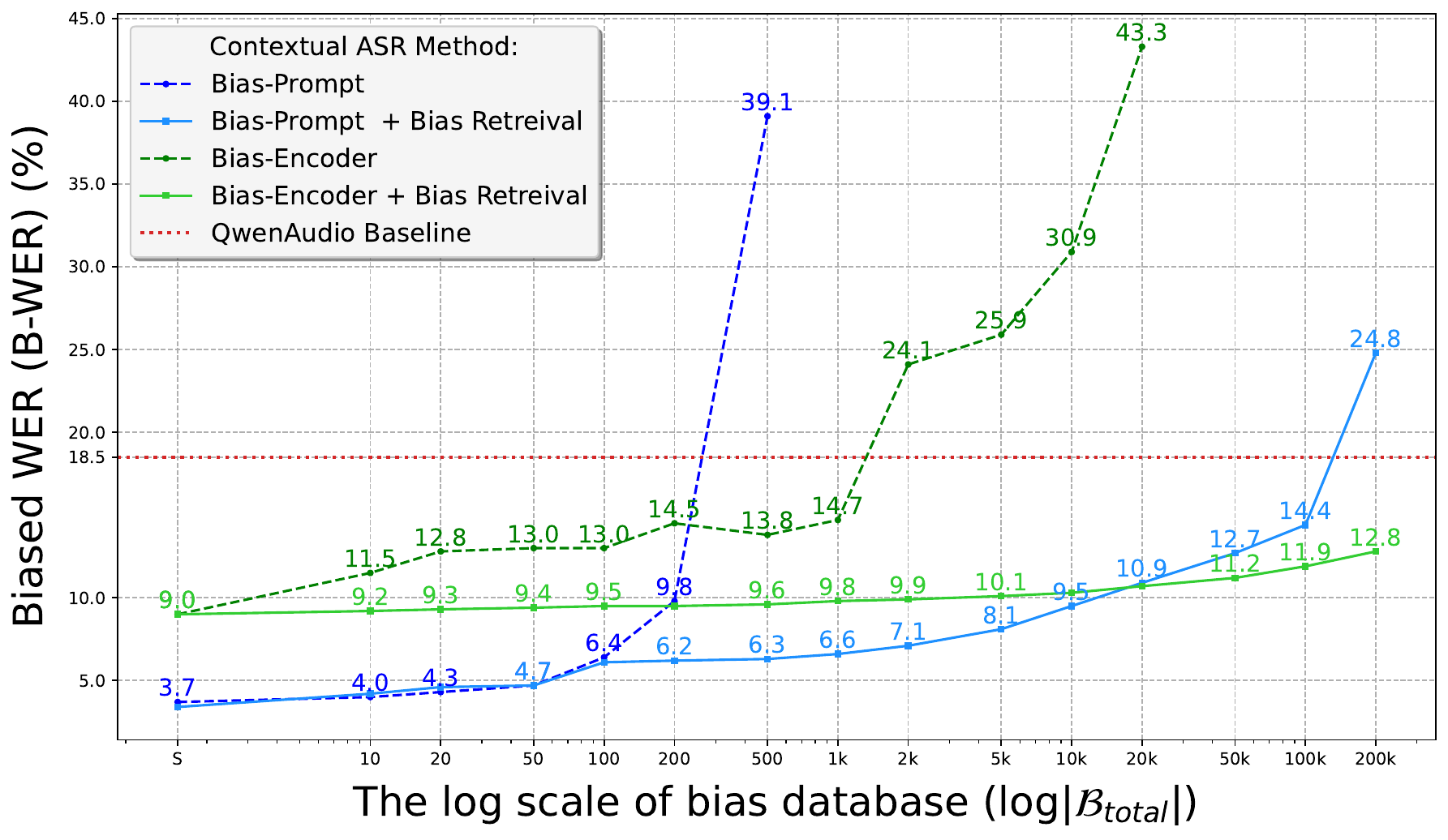}
\caption{The B-WER~(\%) performance across different bias database~($\mathcal{B}_{\text{total}}$) scales from 10 to 200k entries.
Specially, $S=|\mathcal{B}_{\text{sentence}}|$).}
\label{fig:scaling}
\end{figure}

For large-scale bias database up to 200k entries, Fig.~\ref{fig:scaling} shows that the B-WER degradation is manageable.
The BiasEncoder method shows relatively stable performance, where B-WER only degrades by 2.9\% (from 9.9\% to 12.8\%) as $|\mathcal{B}_{\text{total}}|$ expands from 2k to 200k~(i.e., scaling up by 100x), as the Prompt method performance deteriorates faster.
The framework achieves real-time efficiency through FAISS-accelerated retrieval: on a single-thread CPU (Intel Xeon Gold 5318Y), it processes 200k entries with 4096-dimensional vectors in 500ms under 7GB RAM, while GPU acceleration (NVIDIA RTX 4090) reduces the latency to 20ms within 8GB VRAM.

\section{Conclusion}

In this paper, we present BR-ASR: a bias retrieval framework that fundamentally resolves the scalability bottleneck in contextual ASR.
Our proposed method has 4 innovations:
(1) Cross-modal contrastive learning aligns speech inputs with textual bias entries via pretrained speech and bias encoders, achieving 93\% retrieval accuracy while pruning 99.99\% of irrelevant candidates at 200k-scale where the performance degradation is only 0.3\%/2.9\% in terms of WER/B-WER for test-other set.
(2) Dynamic curriculum learning suppresses homophone errors by 30.6\% absolute recall through progressive hard sample scheduling.
Evaluations on LibriSpeech test-clean/-other establish new state-of-the-art performance: 1.2\%/2.8\% WER and 2.8\%/7.1\% B-WER when N=2000.
(3) It demonstrates industrial-grade efficiency with 20ms latency when querying 200k candidates.
(4) BR-ASR enables plug-and-play integration with diverse systems, by decoupling bias retrieval from ASR decoding.

\section{Acknowledgements}

This work was supported in part by China NSFC projects under Grants 62122050 and 62071288, in part by Shanghai Municipal Science and Technology Commission Project under Grant 2021SHZDZX0102, and in part by Ant Group and Ant Group Research Intern Program.

\bibliographystyle{IEEEtran}
\bibliography{mybib}

\end{document}